# Transient Energy and Heat Transport in Metals: Effect of the Discrete Character of the Lattice

Y. Ezzahri,[1*] K. Joulain,[1] and A. Shakouri[2]

1: Institut Pprime, CNRS-Université de Poitiers-ENSMA, Département Fluides, Thermique, Combustion, ENSIP-Bâtiment de mécanique, 40, avenue du Recteur Pineau, F 86022 Poitiers, Cedex, France.
2: Department of Electrical Engineering, University of California Santa Cruz, California 95064 USA
*: younes.ezzahri@univ-poitiers.fr

## ABSTRACT

A recently developed Shastry's formalism for energy transport is used to analyze the temporal and spatial behaviors of the energy and heat transport in metals under delta function excitation at the surface. Comparison with Cattaneo's model is performed. Both models show the transition between nonthermal (ballistic) and thermal (ballistic-diffusive) regimes. Furthermore, because the new model considers the discrete character of the lattice, it highlights some new phenomena such as damped oscillations in the energy transport both in time and space. The energy relaxation of the conduction band electrons in metals is considered to be governed by the electron-phonon scattering, and the scattering time is taken to be averaged over the Fermi surface. Using the new formalism, one can quantify the transfer from nonthermal modes to thermal ones as energy propagates in the material and it is transformed into heat. While the thermal contribution shows a wave-front and an almost exponentially decaying behavior with time, the nonthermal part shows a wave-front and a damped oscillating behavior. Two superimposed oscillations are identified; a fast oscillation that is attributed to the nonthermal nature of energy transport at very short time scales and a slow oscillation that describes the nature of the transition from the nonthermal regime to the thermal regime of energy transport.

# NOMENCLATURE

*a*: lattice constant (Å).
$C_e$: electronic specific heat per unit volume (J.m$^{-3}$.K$^{-1}$).
$D_e$: electronic thermal diffusivity (m²/s).
g: electronic density of states.
H: Heaviside step function.
$I_0$: zero order modified Bessel function of the first kind.
$I_1$: first order modified Bessel function of the first kind.
$k_B$: Boltzmann constant (J/K).
P: input power (W/m$^3$).
q: electronic wave vector (m$^{-1}$).
t: time (s).
T: temperature (K).
U: speed of the heat pulse (m/s).
$v_F$: Fermi velocity (m/s).
x: one dimensional space coordinate (m).
y: nondimensional space coordinate.
Z: thermoelectric figure of merit (K$^{-1}$).

**Greek symbols**

$\beta_e$: electronic thermal conductivity (W/m/K).
$\delta$: Dirac delta function.
$\tau_F$: average relaxation time of conduction band electrons over the Fermi surface of a metal (fs).
$\omega$: angular frequency (rd/s).
$\eta$: nondimensional time.
$\delta K$: energy density (J/m$^3$).
$\varphi_e$: heat flux rate of electrons in the conduction band of the metal (W/m²).
$\xi$: charge-energy modes coupling factor.
$\varepsilon_F$: Fermi energy (J).
$\gamma$: linearity coefficient in $C_e$ (J.m$^{-3}$.K$^{-2}$).

# 1. INTRODUCTION

Development of high-power short-pulse laser sources with a pulse width in the sub-ps range has provided an opportunity to study the propagation of energy and heat at very short time scales; it has also created many applications in thin film analysis or in material processing. Several fundamental physical phenomena in condensed matter such as electronic transitions in semiconductors [1, 2], electron-phonon coupling in metals [3-7], and electron dynamics in semiconductor superlattices [8, 9] have been studied. Shorter laser pulses in the sub-fs range have recently provided the opportunity to explore electronic interactions within the atom itself [10].

Energy transport during short-pulse laser heating of solid materials is an important phenomenon that needs to be fully understood to better control the abundant applications in which short-pulse laser sources are used. The question of energy and heat transport mechanisms at short time and length scales is the basis of numerous theoretical and experimental papers. From a microscopic point of view, energy deposits into and propagates through a material in different ways, depending on the excitation, the structure of the material, and the nature of the energy carriers. At short times and length scales, Fourier's law becomes invalid and many Non-Fourier heat conduction models have been developed to overcome problems associated with the Fourier model (e.g. infinite speed of propagation of heat) [11-14]. Most importantly, the distinction between diffusive and non-diffusive (ballistic) regimes of energy transport becomes very relevant at these short time and length scales [15, 16]. One of the remarkable experimental evidence of ballistic energy transport at small length scales has recently been reported [16].

B. S. Shastry has recently developed a new formalism based on linear theory to describe coupled charge and energy transport in solids [17]. The formalism is general and gives a set of equations for the electro-thermal transport coefficients in the frequency-wave vector domain. One of the most important results of this formalism is the introduction of new response functions describing the change in energy density, charge density, and the currents arising from the input excitation (coefficients $M_1$, $M_2$, $N_1$ and $N_2$ as defined in the reference article [17]). Among these response functions, $N_2$ is of particular interest since this new function gives a measure of the change in the energy density and hence the temperature at various points in the system in response to the applied excitation at the top free surface of the system, and as such represents the energy (heat) Green's function of the system.

In previous works [18, 19], we have analyzed the transient energy and heat transport at the top free surface of a metal occurring after application of a delta-function heating in the frame work of the above mentioned Shastry's formalism. We distinguished between a ballistic and a diffusive contribution to the total energy density. The decomposition we made is based on the physical picture in the momentum space where highly energetic electrons are accelerated ballistically and undergo reflections on the boundaries of the First Brillouin Zone (FBZ) as we shall review later in the discussion section. The motivation behind the current paper is to extend this work and describe the energy density variation as a function of the spatial coordinate. When looking at the physical picture from the real space point of view, it is more appropriate to speak of nonthermal and thermal contributions to the energy density rather than ballistic and diffusive contributions, respectively. We will see that in the nonthermal regime, energy oscillates and one should not really assign a temperature to it. On the other hand, in the thermal regime, heat propagates first ballistically then diffusively. As we have done earlier [18, 19], a detailed comparison with Cattaneo's model [20] is presented both in time and space domains. A short discussion of the frequency dependence of the energy and heat transport at the top free surface of the metal is also presented in the frame work of both models.

The paper is organized as follows; in the first section, we present the theory of energy transport due to a pulsed perturbation at the top free surface of a metal by pointing out the similarities and differences between Shastry's and Catteneo's models. In the second section, the calculation results will be presented by first analyzing the transient energy transport at the top free surface of the metal in both time and frequency domains. Then we will extend the analysis and describe the evolution of the energy density as the observation point is moved through the medium far away from the excitation location at the surface.

## 2. THEORY

### 2.1 Shastry's model

We follow the calculations performed in recent papers [18, 19], and the starting point of our analysis will be the Shastry-Green function $N_2^S$ in the decoupled limit for metals. According to Shastry's work [17], the coupling factor $\xi$ between charge and energy modes can be expressed using the high frequency value of the thermoelectric figure-of-merit $Z^*T$:

$$\xi = \frac{Z^*T}{Z^*T+1} \quad (1)$$

Here $Z^*$ is the high frequency limit of the Seebeck coefficient square times the electrical conductivity divided by the thermal conductivity. It is well known, however, that metals are very poor thermoelectric materials with a very low $ZT$ [21]. The decoupled limit is thus justified. By turning off the coupling between the charge and energy modes ($\xi=0$), $N_2^S$ can be expressed as:

$$N_2^S(\omega,q) = \frac{-i+\tau_q\omega}{\omega+i\tau_q\omega^2-iD_eq^2} \quad (2)$$

We should note here that Eq (2) is given for an arbitrary applied power excitation function $P(t)$ at the top free surface of a metal once it is described as a function of its Fourier components [17]. $\omega$ is the angular frequency, $q$ is the electron wave vector, $D_e$ is the electronic thermal diffusivity and $\tau_q$ is the total electron scattering time, which, in general, is a function of $q$. Remarkably, in the case of a $q$-independent $\tau_q$, the form of $N_2^S$ in the frequency-wave vector domain, resembles the expression of the energy density change at the top free surface of the metal, one would have derived using the hyperbolic model of Cattaneo in the case of a delta power excitation applied to the surface. We will get back to Cattaneo's model more in detail in the next section.

In the following, we consider a one dimensional energy transport problem, in which case we assume the top metal free surface being excited by an input perturbation pulse like a laser pulse of power $P(t)$. The one dimensional approximation is reasonable at short time scales considering the ratio of the size of the laser pulse spot to the diffusion length in couple of nanoseconds. The delta function excitation is valid since the optical penetration depth is small. The latter quantity depends on the wavelength of the laser, but it is less than *10nm* over a large range of wavelengths for most metals [22]. The optical penetration depth is on the order of *1nm* if very short wavelengths are used (e.g. UV with a frequency lower than the Plasmon frequency of the corresponding metal to avoid any resonance phenomenon) [22]. This is very useful since it justifies the assumption for surface excitation. Figure 1 shows a schematic of the metal under excitation, where we assume a laser pulse to excite the top free surface (x=0).

After excitation of the top metal free surface with an input power *P(t)*, the change in the energy density propagation in the material in the frequency-wave vector domain $\overline{\widetilde{\delta K}}_S(\omega, q)$ can be expressed as [17]:

$$\overline{\widetilde{\delta K}}_S(\omega, q) = N_2^S(\omega, q) \times \overline{P}(\omega) \quad (3)$$

where $\overline{P}(\omega)$ is the Fourier transform of *P(t)* in the frequency domain. The change in the energy density at any point of the metal as function of time is obtained by a double inverse Fourier transforms with respect to $\omega$ and *q*:

$$\delta K_S(t, x) = \frac{1}{(2\pi)^2} \int_{-\pi/a}^{\pi/a} \left[ \int_{-\infty}^{+\infty} N_2^S(\omega, q) \overline{P}(\omega) e^{i\omega t} d\omega \right] e^{iqx} dq \quad (4)$$

The integration over *q* is taken over the First Brillouin Zone (FBZ) in the one dimensional case, where "*a*" refers to the lattice constant of the metal. The power source *P(t)* can be of any form, but we will limit our study to the ideal case of a Dirac delta function $P(t) = P_0 \times \delta(t)$ in order to capture the intrinsic evolution of the energy density *δK(t, x)* as a function of time at different locations of the metal, i.e. *the time-space behavior of the energy density Green's function*.

The integral within the square brackets in Eq (4) can be analytically calculated using the residue theorem. The integrand of this integral has two single poles that lie in the upper complex half plane. These poles are given by:

$$\omega_{\pm} = i \frac{1 \pm \sqrt{1 - 4 D_e \tau_q q^2}}{2\tau_q} \quad (5)$$

A straightforward calculation of the residues at these two single poles leads to:

$$\int_{-\infty}^{+\infty} N_2^S(\omega, q) e^{i\omega t} d\omega = 2\pi e^{-\frac{t}{2\tau_q}} \left[ ch\left(\frac{R_q t}{2\tau_q}\right) + \frac{sh\left(\frac{R_q t}{2\tau_q}\right)}{R_q} \right] = 2\pi N_2^S(t, q)$$

$$\text{with } R_q = \sqrt{1 - 4 D_e \tau_q q^2} \quad (6)$$

Based on this complex analysis calculation, we can demonstrate also that $N_2^S$ is a causal function ( $N_2^S(t<0, q) = 0 \ \forall q$ ).

Equation (4) can then be re-expressed as:

$$\delta K_S(t, x) = \frac{P_0}{2\pi} \int_{-\pi/a}^{\pi/a} e^{-\frac{t}{2\tau_q}} \left[ ch\left(\frac{R_q t}{2\tau_q}\right) + \frac{sh\left(\frac{R_q t}{2\tau_q}\right)}{R_q} \right] e^{iqx} dq \quad (7)$$

Up to now, the equations are completely general regardless the dependence of the relaxation time on the wave vector. Electron energy relaxation in the conduction band of metals is governed by the electron-phonon scattering processes [23]. The electron-electron scattering processes are generally faster but they don't change the total energy of the electron gas. In most metals the transport of energy by phonons themselves can be neglected. Using the fact that electrons and phonons in a metal can be characterized by different temperatures when their respective distributions reach a quasi-equilibrium state, it has been shown that scattering of electrons by phonons can be either elastic or inelastic, and the relaxation time is inversely proportional to the lattice temperature [4, 24-26]. In the following, we consider the case of a constant relaxation time $\tau_q = \tau_F$, which we consider to be the average scattering time of electrons over the Fermi surface of the metal. Eq (7) can be split into two parts, depending on the sign of the argument of $R_q$. We can write it down as:

$$\begin{cases} \delta K_S(t,x) = \delta K_S^<(t,x) + \delta K_S^>(t,x) \\[2mm] \delta K_S^<(t,x) = \dfrac{P_0}{\pi} e^{-\dfrac{t}{2\tau_F}} \displaystyle\int_0^{q_0} \left[ ch\left(\dfrac{R_q t}{2\tau_F}\right) + \dfrac{sh\left(\dfrac{R_q t}{2\tau_F}\right)}{R_q} \right] \cos(qx)\,dq \quad (a) \\[4mm] \delta K_S^>(t,x) = \dfrac{P_0}{\pi} e^{-\dfrac{t}{2\tau_F}} \displaystyle\int_{q_0}^{q_m} \left[ \cos\left(\dfrac{\overline{R_q} t}{2\tau_F}\right) + \dfrac{\sin\left(\dfrac{\overline{R_q} t}{2\tau_F}\right)}{\overline{R_q}} \right] \cos(qx)\,dq \quad (b) \\[4mm] q_m = \dfrac{\pi}{a};\ q_0 = \dfrac{1}{2\sqrt{D_e \tau_F}} = \dfrac{1}{2U\tau_F} \text{ and } \overline{R_q} = \sqrt{4D_e \tau_F q^2 - 1} \end{cases} \qquad (8)$$

Eq (8) was simplified using the fact that $N_2^S(t,q)$ is an even function of $q$, only the integral of the real part remains, while the integral of the imaginary part vanishes. $U = \sqrt{D_e/\tau_F}$ represents the propagation speed of the energy pulse. We should note here that this decomposition of the integral in Eq (7) is allowed because both of the integrals in Eq (8) are finite integrals and that there are no singularities of the integrand at the separation wave vector $q_0$. Indeed, it is very easy to show that:

$$\lim_{q \to \overline{q_0}} \left[ ch\left(\frac{R_q t}{2\tau_F}\right) + \frac{sh\left(\frac{R_q t}{2\tau_F}\right)}{R_q} \right] \cos(qx) = \lim_{q \to q_0^+} \left[ \cos\left(\frac{\overline{R_q} t}{2\tau_F}\right) + \frac{\sin\left(\frac{\overline{R_q} t}{2\tau_F}\right)}{\overline{R_q}} \right] \cos(qx)$$

$$= \left(1 + \frac{t}{2\tau_F}\right) \cos\left(\frac{x}{2U\tau_F}\right), \forall t \geq 0 \text{ and } \forall x \quad (9)$$

The first part in Eq (8) describes the diffusive-hyperbolic contribution, it is exponentially decaying as function of time and it has the energy front propagation as we shall see later. One could assign a "temperature" to this part of the energy propagation and as such, this part constitutes the thermal contribution to the energy density. On the other hand, the second part describes the nonthermal behavior. It is interesting to note the damped oscillating character of the integrand in Eq 8(b) which can be seen after integration over $q$. Since the energy propagation will oscillate (moving back and forth), it does not make sense to assign a "temperature" to this aspect of energy transport.

When dealing with energy and heat transfer problems, it is preferable to write down the representative equations using nondimensional variables. Let us put $\eta = t/\tau_F$ and $y = x/U\tau_F$ the nondimensional time and space coordinate, respectively. Eq (8) can then be rewritten as:

$$\begin{cases} \delta K_S(\eta, y) = \delta K_S^<(\eta, y) + \delta K_S^>(\eta, y) \\ \delta K_S^<(\eta, y) = \dfrac{P_0}{\pi U \tau_F} e^{-\eta/2} \int_0^{1/2} \left[ ch\left(R_Q \dfrac{\eta}{2}\right) + \dfrac{sh\left(R_Q \dfrac{\eta}{2}\right)}{R_Q} \right] \cos(Qy)\, dQ \quad (a) \\ \delta K_S^>(\eta, y) = \dfrac{P_0}{\pi U \tau_F} e^{-\eta/2} \int_{1/2}^{Q_m} \left[ \cos\left(\overline{R_q} \dfrac{\eta}{2}\right) + \dfrac{\sin\left(\overline{R_q} \dfrac{\eta}{2}\right)}{\overline{R_q}} \right] \cos(Qy)\, dQ \quad (b) \\ Q_m = U \tau_F q_m \text{ and } R_Q = \sqrt{1 - 4Q^2} = i\overline{R_Q} \end{cases} \quad (10)$$

In the next section, we will show that a similar decomposition in the time-space domain can be performed using the hyperbolic model of Cattaneo.

## 2.2 Cattaneo's model

Cattaneo's model is one of the simplest, most powerful and widely used Non-Fourier models that have been developed to overcome the shortcoming or paradox of an infinite propagation speed of thermal signals which is a characteristic of Fourier's parabolic diffusion equation. According to Fourier's model, the energy density Green's function using nondimensional variables $\eta$ and $y$ is given by:

$$\delta K_F(\eta, y) = \dfrac{P_0}{2 U \tau_F \sqrt{\pi \eta}} e^{-\dfrac{y^2}{4\eta}} \quad (11)$$

As we shall see below, the resulting energy equation based on Cattaneo's model is hyperbolic and is often used to study the temperature and heat flux fields in metallic and dielectric materials with submicrometer thickness as well as in tightly packed microelectronic devices with imposed boundary conditions [27].

Our starting point here is the one dimensional Cattaneo's equation applied to electrons in the conduction band of a metal. This equation relates the heat flux rate of electrons to their temperature gradient and is given by [20]:

$$\tau_F \frac{\partial \varphi_e}{\partial t} + \varphi_e = -\beta_e(T_e)\frac{\partial T_e}{\partial x} \quad (12)$$

where $\beta_e$ represents the thermal conductivity of the electrons. To this equation we add the energy conservation equation for a metal, which is given by:

$$\frac{\partial [\delta K_C]}{\partial t} + \frac{\partial \varphi_e}{\partial x} = P(t,x) \quad (13)$$

where $P(t,x)$ represents the input power applied to the top free surface of the metal. On the other hand, the energy density of electrons is related to their temperature via the equation [23, 24]:

$$\delta K_C(t,x) = \frac{1}{2}C_e(T_e)T_e(t,x) \quad (14)$$

where $C_e$ is the temperature-dependent specific heat per unit volume of the electronic system. $C_e(T_e) = \pi^2/3 \, k_B^2 g(\varepsilon_F)T_e = \gamma T_e$ where $g(\varepsilon_F)$ is the electronic density of states at the Fermi energy $\varepsilon_F$. For gold, the linearity coefficient is given by $\gamma=66$ J/m$^3$/K$^2$ [25]. Combination of Eqs (12-14) allows us to write the energy density equation as:

$$\tau_F \frac{\partial^2[\delta K_C]}{\partial t^2} + \frac{\partial[\delta K_C]}{\partial t} - D_e\frac{\partial^2[\delta K_C]}{\partial x^2} = \tau_F \frac{\partial P}{\partial t} + P \quad (15)$$

As we have done in Shastry's model, we assume the metal to be excited at its top free surface by a laser delta pulse of power $P(t,x) = P_0 \times \delta(t,x)$.

The solution of the hyperbolic equation (15) either for an infinite or semi-infinite domain, has been extensively investigated using different methods (see, for example, references [27-30]). In this particular case of time and space Dirac delta excitation, we can easily demonstrate that the solution for a semi-infinite bounded body is just a factor of two times the solution for an infinite unbounded body. This can be derived using the symmetry property of the energy density with respect to x. Although, our case study lies within the configuration of a semi-infinite domain, we continue to use the solution of an infinite body throughout the whole paper which simplifies the analysis. As mentioned in reference [27], when $P(t,x) = P_0 \times \delta(t,x)$ is viewed as a volumetric heat source, one can consider the x=0 surface to be insulated. The solution of Eq (15) can be written using nondimensional variables $\eta$ and $y$ as:

$$\delta K_C(\eta, y) = \frac{P_0}{2U\tau_F} e^{-\frac{\eta}{2}} \left\{ \begin{array}{l} I_0[\chi(\eta,y)]\delta(\eta-|y|) \\ + \frac{1}{2}\left[ I_0[\chi(\eta,y)] + \frac{\eta}{2\chi(\eta,y)} I_1[\chi(\eta,y)] \right] H(\eta-|y|) \end{array} \right\}$$

with $\chi(\eta, y) = \frac{1}{2}\sqrt{\eta^2 - y^2}$ (16)

Where $I_0$, $I_1$ and $H$ are the zero order and first order modified Bessel functions of the first kind, and Heaviside function, respectively. This solution is characterized by two main terms. The presence of the Dirac delta function in the first term is a reproduction of the initial pulse source damped by the exponential factor and propagating with speed U. The second term which includes the Heaviside function corresponds to the wake, which is a manifestation of the causality requirement so that a perturbation front propagates with a speed U, while the positions beyond the front ($y>\eta$) are unaltered by the action of the energy pulse [28, 29]. For sufficiently long times ($\eta>>y$), it is this term which yields the usual diffusion solution in Eq (11) as we shall see later in the discussion section.

We can obtain the same solution in Eq (16) by proceeding in a similar way as in Shastry's model by first transforming the whole energy equation (15) using a double non-unitary Fourier transform with respect to time and space. In fact, by doing so, Eq (15) becomes:

$$\left(-\tau_F \omega^2 + i\omega + D_e q^2\right) \overline{\widetilde{\delta K}}_C = \left(1 + i\tau_F \omega\right) P_0 \quad (17)$$

From which we can easily extract the expression of the energy density in the frequency-wave-vector domain:

$$\overline{\widetilde{\delta K}}_C(\omega, q) = \frac{1 + i\tau_F \omega}{-\tau_F \omega^2 + i\omega + D_e q^2} P_0 = \frac{-i + \tau_F \omega}{\omega + i\tau_F \omega^2 - iD_e q^2} P_0 \quad (18)$$

It is very interesting to note the resemblance between Eqs (2) and (18). Both Shastry and Cattaneo models, give similar expressions to the ratio $\dfrac{\overline{\widetilde{\delta K}}(\omega, q)}{P_0}$.

The change in the energy density Green's function at any point of the metal as function of time is obtained by a double inverse Fourier transforms of $\overline{\widetilde{\delta K}}_C(\omega, q)$ with respect to $\omega$ and $q$, the difference with respect to Shastry's model, is that $q$ can vary between $-\infty$ and $+\infty$:

$$\delta K_C(t, x) = \frac{P_0}{2\pi} e^{-\frac{t}{2\tau_F}} \int_{-\infty}^{+\infty} \left[ ch\left(\frac{R_q t}{2\tau_F}\right) + \frac{sh\left(\dfrac{R_q t}{2\tau_F}\right)}{R_q} \right] e^{iqx} dq \quad (19)$$

This equation can also be written in another form by noting the relation between the two terms of the integrand:

$$\delta K_C(t, x) = \frac{P_0}{2\pi} e^{-\frac{t}{2\tau_F}} \left\{ \int_{-\infty}^{+\infty} \frac{sh\left(\dfrac{R_q t}{2\tau_F}\right)}{R_q} e^{iqx} dq + 2\tau_F \frac{\partial}{\partial t} \left[ \int_{-\infty}^{+\infty} \frac{sh\left(\dfrac{R_q t}{2\tau_F}\right)}{R_q} e^{iqx} dq \right] \right\} \quad (20)$$

Using complex analysis [28], we can easily show that:

$$\int_{-\infty}^{+\infty} \frac{sh\left(\dfrac{R_q t}{2\tau_F}\right)}{R_q} e^{iqx}\, dq = \frac{\pi}{2U\tau_F} I_0\left[\frac{1}{2}\sqrt{\left(\frac{t}{\tau_F}\right)^2 - \left(\frac{x}{U\tau_F}\right)^2}\right] H(Ut - |x|) \quad (21)$$

Introducing Eq (21) into Eq (20) will result after simplification in:

$$\delta K_C(t,x) = \frac{P_0}{2U\tau_F} e^{-\frac{t}{2\tau_F}} \left\{ \begin{array}{l} U\tau_F I_0[\chi(t,x)]\delta(Ut - |x|) \\ + \dfrac{1}{2}\left[I_0[\chi(t,x)] + \dfrac{t/\tau_F}{2\chi(t,x)} I_1[\chi(t,x)]\right] H(Ut - |x|) \end{array} \right\} \quad (22)$$

with $\chi(t,x) = \dfrac{1}{2}\sqrt{\left(\dfrac{t}{\tau_F}\right)^2 - \left(\dfrac{x}{U\tau_F}\right)^2}$

This is exactly the classical solution in Eq (16) by replacing dimensional variables $t$ and $x$ by nondimensional ones $\eta$ and $y$.

We have thus derived similar expressions for the ratio $\dfrac{\widetilde{\delta K}(\omega, q)}{P_0}$ from both Shastry and Cattaneo models, which means we can perform the same time-space decomposition of the electronic energy density in a metal using Cattaneo's model. The causality requirement which is described mathematically using Heaviside function $H$ in Eq (21) is a direct consequence of application of Jordan's lemma in complex analysis for the case of Cattaneo's model. Based on this, we can write the left hand side of Eq (21) as:

$$\int_{-\infty}^{+\infty} \frac{sh\left(\dfrac{R_q t}{2\tau_q}\right)}{R_q} e^{iqx} dq = \int_{-\infty}^{+\infty} \frac{sh\left(\dfrac{R_q t}{2\tau_q}\right)}{R_q} e^{iqx} dq \times H(Ut - |x|)$$

$$= 2\left[\int_0^{q_0} \frac{sh\left(\dfrac{R_q t}{2\tau_q}\right)}{R_q} \cos(qx)\, dq + \int_{q_0}^{+\infty} \frac{sh\left(\dfrac{R_q t}{2\tau_q}\right)}{R_q} \cos(qx)\, dq\right] H(Ut - |x|) \quad (23)$$

To write the second line in Eq (23), we used the fact that the integrand is an even function of $q$. By using the decomposition in Eq (23) with nondimensional variables $\eta$ and $y$, and arranging the terms, Eq (20) can be split and expressed as:

$$\begin{cases} \delta K_C(\eta, y) = \delta K_C^<(\eta, y) + \delta K_C^>(\eta, y) \\ \\ \delta K_C^<(\eta, y) = \dfrac{P_0}{\pi U \tau_F} e^{-\frac{\eta}{2}} \left\{ \begin{array}{l} 2\delta(\eta - |y|) \displaystyle\int_0^{1/2} \dfrac{sh\left(R_Q \frac{\eta}{2}\right)}{R_Q} \cos(Qy)\, dQ \\ \\ + H(\eta - |y|) \displaystyle\int_0^{1/2} \left[ ch\left(R_Q \frac{\eta}{2}\right) + \dfrac{sh\left(R_Q \frac{\eta}{2}\right)}{R_Q} \right] \cos(Qy)\, dQ \end{array} \right\} \quad (a) \\ \\ \delta K_C^>(\eta, y) = \dfrac{P_0}{\pi U \tau_F} e^{-\frac{\eta}{2}} \left\{ \begin{array}{l} 2\delta(\eta - |y|) \displaystyle\int_{1/2}^{+\infty} \dfrac{\sin\left(\overline{R_Q} \frac{\eta}{2}\right)}{\overline{R_Q}} \cos(Qy)\, dQ \\ \\ + H(\eta - |y|) \displaystyle\int_{1/2}^{+\infty} \left[ \cos\left(\overline{R_Q} \frac{\eta}{2}\right) + \dfrac{\sin\left(\overline{R_Q} \frac{\eta}{2}\right)}{\overline{R_Q}} \right] \cos(Qy)\, dQ \end{array} \right\} \quad (b) \\ \\ R_Q = \sqrt{1 - 4Q^2} = i\overline{R_Q} \end{cases} \quad (24)$$

We can see that, by following this procedure, the energy density Green's function in the frame work of Cattaneo's model is expressed as the sum of two contributions; the ballistic-diffusive (thermal) contribution $\delta K_C^<(\eta, y)$ and the oscillatory (nonthermal) contribution $\delta K_C^>(\eta, y)$.

In analyzing Shastry's model, we have not taken into consideration any causality requirement, since it does not appear automatically in the mathematical procedure. This is in contrary to Cattaneo's model as mentioned above. Not having the causality requirement will produce anomalies in the behavior of the energy density in the time-space domain. In order to overcome these anomalies, one can look to Shastry's model from another angle. Because of the finite wave-vector bounds in Shatsry's model, we can mathematically analyze this model differently by writing:

$$\delta K_S(t,x) = \frac{P_0}{2\pi} \int_{-q_m}^{q_m} N_2^S(t,q) e^{iqx} dq = \frac{P_0}{2\pi} \int_{-\infty}^{\infty} rect_{q_m}(q) N_2^S(t,q) e^{iqx} dq$$

$$= \frac{1}{2\pi} \int_{-\infty}^{\infty} rect_{q_m}(q) e^{iqx} dq \otimes \delta K_C(t,x) \quad (25)$$

Where $rect_{q_m}(q) = H(q+q_m) - H(q-q_m)$ is the *"rectangular function"* [31] whose inverse Fourier transverse is the kernel function $\frac{q_m}{\pi} Sinc(q_m x)$ with $sinc(x) = \sin(x)/x$ being the *"sinc function"*[31]. The inconvenience of following this procedure in the analysis is the difficulty to separate the effect of the cut-off in the wave-vector domain on the behavior of the energy density Green's function for electrons in the metal. However, it constitutes a robust argument to add the causality requirement to Shastry's model to meet physical requirements even though this is not mathematically rigorous. On the basis of this idea, the expressions of both the ballistic-diffusive (thermal) and oscillatory (nonthermal) contributions in Shastry's model as given by Eqs (10a) and (10b) have to be multiplied by $H(\eta - y)$, respectively, in the reminder of our discussion.

We close this section on the theory of both formalisms, by discussing the frequency dependence of the energy density Green's function.

Using the same expression of the separation wave-vector $q_0$, the time decomposition can be translated to decomposition in the frequency domain. For both Shastry and Cattaneo models, the Green's function $N_2^S$ is directly integrated in the wave-vector domain.

$$\begin{cases} \delta K_{S,C}(\omega,0) = \delta K_{S,C}^<(\omega,0) + \delta K_{S,C}^>(\omega,0) \\ \delta K_{S,C}^<(\omega,0) = \dfrac{P_0}{\pi} \int_0^{q_0} N_2^S(\omega,q)\, dq \quad (a) \\ \delta K_S^>(\omega,0) = \dfrac{P_0}{\pi} \int_{q_0}^{q_m} N_2^S(\omega,q)\, dq \quad (b) \\ \delta K_C^>(\omega,0) = \dfrac{P_0}{\pi} \int_{q_0}^{\infty} N_2^S(\omega,q)\, dq \quad (c) \end{cases} \quad (26)$$

The thermal and nonthermal contributions to the total energy density Green's function variation at the top free surface of a metal, are given in the frequency domain by the following expressions for both Shastry [Eqs (27)] and Cattaneo [Eqs (28)] models:

$$\begin{cases} \delta K_S^<(\omega,0) = \dfrac{P_0}{\pi} \sqrt{\dfrac{1+i\tau_F\omega}{iD_e\omega}} \mathrm{Atan}\left(\dfrac{1}{2\sqrt{i\tau_F\omega - \tau_F^2\omega^2}}\right) \quad (a) \\ \delta K_S^>(\omega,0) = \dfrac{P_0}{\pi} \sqrt{\dfrac{1+i\tau_F\omega}{iD_e\omega}} \left[ \mathrm{Atan}\left(\dfrac{\pi}{a}\sqrt{\dfrac{D_e}{i\omega - \tau_F\omega^2}}\right) - \mathrm{Atan}\left(\dfrac{1}{2\sqrt{i\tau_F\omega - \tau_F^2\omega^2}}\right) \right] \quad (b) \end{cases} \quad (27)$$

$$\begin{cases} \delta K_C^<(\omega,0) = \dfrac{P_0}{\pi} \sqrt{\dfrac{1+i\tau_F\omega}{iD_e\omega}} \mathrm{Atan}\left(\dfrac{1}{2\sqrt{i\tau_F\omega - \tau_F^2\omega^2}}\right) \quad (a) \\ \delta K_C^>(\omega,0) = \dfrac{P_0}{\pi} \sqrt{\dfrac{1+i\tau_F\omega}{iD_e\omega}} \left[ \dfrac{\pi}{2} - \mathrm{Atan}\left(\dfrac{1}{2\sqrt{i\tau_F\omega - \tau_F^2\omega^2}}\right) \right] \quad (b) \end{cases} \quad (28)$$

In these equations, (a) and (b) describe the thermal (ballistic-diffusive) and the non-thermal (ballistic) contributions, respectively. It is straightforward to deduce the total energy density Green's function in the frequency domain at the top free surface of a metal:

$$\begin{cases} \delta K_S(\omega,0) = \dfrac{P_0}{\pi}\sqrt{\dfrac{1+i\tau_F\omega}{iD_e\omega}}\,\mathrm{Atan}\!\left(\dfrac{\pi}{a}\sqrt{\dfrac{D_e}{i\omega-\tau_F\omega^2}}\right) & (a) \\ \delta K_C(\omega,0) = \dfrac{P_0}{2}\sqrt{\dfrac{1+i\tau_F\omega}{iD_e\omega}} & (b) \end{cases} \quad (29)$$

One can easily check that Eqs 27(a), 28(a), 29(a) and 29(b) tend to a Fourier type equation at low frequencies:

$$\delta K_S^<(\omega\to 0,0) = \delta K_C^<(\omega\to 0,0) = \dfrac{P_0}{2}\sqrt{\dfrac{1}{iD_e\omega}} = \delta K_F(\omega,0) \quad (30)$$

## 3. RESULTS AND DISCUSSION

We start our discussion section by first analyzing both time and frequency behaviors of the change in the electron energy density Green's function at the top free surface of the metal, then we will expand the analysis to look for the behavior at different locations $y$ in the direction of the energy transport through the medium.

At electron temperatures, $T_e$, smaller than the Fermi temperature, $T_F$, the electron thermal conductivity is given by $\beta_e = C_e v_F^2 \tau_F / 3$ [23], where $v_F$ is the Fermi velocity, and $\tau_F$ is the scattering relaxation time of electrons averaged over the Fermi surface of the metal. The thermal diffusivity of the electronic system takes, then, a simple expression $D_e = \dfrac{\beta_e}{C_e} = \dfrac{v_F^2 \tau_F}{3}$ whose temperature dependence is due to $\tau_F$. This is under the assumption that $v_F$ is temperature independent.

In Figs 2, we have plotted the temporal behavior of the different contributions to the energy density Green's function in Shastry's model at the top free surface of two metals, gold and aluminum, at room temperature $T=300K,$ as given by Eqs (8a) and (8b). More specifically the vertical axis in Figs 2(a) and 2(b) represents the quantity $\delta K_S(t,0)/P_0$ which has the unit of the absorption coefficient ($m^{-1}$). The higher is this quantity the higher is the energy density and hence the temperature at the top free surface of the metal. The values of the scattering relaxation time $\tau_F$ are estimated based on the values of the electrical resistivity using Drude theory [23]. Table 1 summarizes these values as well as the values of the Fermi velocities and lattice constants of gold and aluminum at room temperature.

The left side of the integral as described by Eq. 8(a) shows a smooth decaying behavior as a function of time which is almost an exponential [Fig 2(a)]. This behavior is characteristic of the diffusive regime of heat transport by electrons. On the other hand, the right side of the integral described by Eq. 8(b) shows an oscillating behavior as a function of time [Fig 2(b)]. Two oscillations can be identified and are damped out exponentially with time. The periods of these oscillations can be estimated using Eq (8b) expressed at the top surface of the metal (x=0) as follows.

If we consider values of the wave-vector $q$ between $q_0$ and $q_m$, we can, with a very good approximation, neglect $1$ in the argument of $\overline{R_q}$, the latter becomes then $\overline{R_q} \approx 2q\sqrt{D_e \tau_F} = 2qU\tau_F$ and Eq (8b) can be written as:

$$\delta K_S^>(t,0) = \frac{P_0}{\pi} e^{-\frac{t}{2\tau_F}} \int_{q_0}^{q_m} \left[ \cos(qUt) + \frac{\sin(qUt)}{2qU\tau_F} \right] dq \quad (31)$$

This integral can be calculated analytically, the result of which is given by:

$$\delta K_S^>(t,0) = \frac{P_0}{\pi} e^{-\frac{t}{2\tau_F}} \left[ \frac{\sin(q_m Ut) - \sin\left(\frac{t}{2\tau_F}\right)}{Ut} + \frac{Si(q_m Ut) - Si\left(\frac{t}{2\tau_F}\right)}{2U\tau_F} \right] \quad (32)$$

Where *Si* represents the *"Sine integral function"* given by:

$$Si(z) = \int_0^z \frac{\sin(x)}{x} dx = \int_0^z \sin c(x) dx \ [31].$$

As we can see in this equation, the result contains two main periods or pseudo-periods, $\theta_F = 2\pi/q_m U = 2a/U = 2\sqrt{3}\frac{a}{v_F}$ and $\theta_S = 4\pi\tau_F$ where $\theta_F$ and $\theta_S$ describe the fast and slow oscillation periods, respectively. $\theta_S$ can be seen as the damped oscillation of the envelope.

The fast oscillation period is function only of the lattice constant of the metal and its Fermi velocity; moreover it is independent of the scattering relaxation time of the electrons in the conduction band. For almost all metals $a \sim 4 Å$ and $v_F$ is of the order of $1.4 \times 10^6$ m/s, a simple application shows then that the period of this oscillation is very small $\theta_F \approx 1 fs$ which makes the fastest oscillation [Fig 2(b)]. On the other hand, the slow oscillation period is proportional to the total scattering relaxation time of electrons, which can be of few tenths of femtoseconds. Furthermore, by changing the temperature, only the scattering relaxation time is affected; the relaxation time $\tau_F$ increases by decreasing the temperature [23, 24, 32]. Since the fast oscillation period is independent of $\tau_F$, this period is also independent of the temperature [18], while the slow oscillation period increases by decreasing the temperature.

The fast oscillating behavior in the energy (heat) transport that results from Shastry's formalism is a consequence of the band cut-off due to the discrete character of the lattice; the oscillations are caused by Bragg reflections of ballistically accelerated electrons at the boundaries of the First Brillouin Zone (FBZ). Based on the physical picture in the momentum space (reciprocal space), these electrons can make many round-trips within the FBZ bouncing back and forth on the boundaries, before they damp out due scattering mechanisms inside this zone. This is characteristic of the nonthermal aspect of the energy density, during which the distribution of electrons is in a nonequilibrium state. Based on the physical picture in the real space, the nonthermal (ballistic) electrons become afterwards thermal (ballistic-diffusive). This is illustrated by the difference of the amplitudes of $\delta K_S^< / P_0$ and $\delta K_S^> / P_0$ in Figs 2(a) and 2(b). At short time scales, energy is mostly in a nonthermal state; the amplitude of the nonthermal contribution is higher than the amplitude of the thermal contribution and as the time goes by, the nonthermal regime transition to a thermal regime. For higher electron energies (high Fermi velocities), the oscillation period is shorter and the number of reflections is increased before nonthermal energy transport damps out to a thermal regime.

This fast oscillating behavior in the energy density modes can be viewed as the analogous of the Bloch oscillations of the electronic charge density in the material subject to a uniform electric field [8, 9]. The period of charge density Bloch oscillations is inversely proportional to the lattice constant, while this period is proportional to the lattice constant for the energy density modes oscillations as shown from the expression of the fast oscillation period $\theta_F$.

While the fast oscillations are attributed to the nonthermal character of energy transport at short time scales, the slow oscillations on the other hand, can be interpreted as a manifestation of the nature of the transition between the nonthermal (ballistic) and thermal (ballistic-diffusive) regimes in energy transport by the electronic system in the conduction band of the metal. Two arguments support this conclusion; *(i)* the amplitude of the slow oscillations as compared to the amplitude of the fast oscillations and *(ii)* the proportionality between the slow oscillation period $\theta_S$ and the relaxation time $\tau_F$. The transition from the nonthermal regime to the thermal regime in energy transport does not happen suddenly, but it occurs in a damped oscillating manner.

In Figs 3(a) we have plotted the thermal contribution to the energy density Green's function at the top free surface of the two metals as calculated based on Shastry's model [Eq 8(a)] as a function of the nondimensional time $\eta$, in comparison to Fourier's diffusion prediction which is simply given by $\delta K_F(\eta,0) = \dfrac{P_0}{2U\tau_F\sqrt{\pi\eta}}$. The temporal behavior of the thermal contribution $\delta K_S^<(\eta,0)$, can be divided into two regimes at short and at long time scales. For $\eta<1$, before any scattering event happens, $\delta K_S^<(\eta,0)$ is almost constant, then it starts decaying exponentially with time up to about $\eta\sim 8$, from where it changes the trend and it starts following a Fourier type energy diffusion law. As it can be seen in Fig 3(a), Shastry and Fourier models converge perfectly at long time scales.

In Fig 3(b), we show separately the temporal behavior of the thermal contribution and the sum of the nonthermal and thermal contributions to the total energy density Green's function at the top free surface of gold at room temperature, as calculated based on Shastry's model [Eqs 8(a), and 8(a+b)], in comparison to Fourier's model. While the nonthermal contribution is the dominant one at short time scales, it becomes insignificant after about $8\tau_F$-$10\tau_F$. After that moment, the total energy is transported diffusively in which case the temporal decay follows a Fourier type law.

As we have mentioned above, the closest Non-Fourier model to Shastry's model is Cattaneo's model [11, 20]. We have shown that both models allow the separation between the nonthermal and the thermal contributions to the total energy density variation as a function of time and space. Both models describe the thermal regime using similar expression [Eqs 10(a) and 24(a)]. On the other hand, because of the continuous character of Cattaneo's model, the upper limit in the integral of Eq 24(b) is infinity, and the nonthermal regime is expected to show a different behavior than Shastry's. As we did before, for values of the wave vector $q$ between $q_0$ and infinity, we neglect $1$ in the argument of $\overline{R_q}$ and we use the approximating expression $\overline{R_q} \approx 2qU\tau_F$. Using this approximation and dropping off the first term in Eq(24b), the latter equation expressed at the top free surface of the metal x=0, can be calculated analytically and the result is given by:

$$\delta K_C^>(t,0) = \frac{P_0}{U} e^{-\frac{t}{2\tau_F}} \left[ \delta(t) + \frac{\frac{\pi}{2} - \sin c\left(\frac{t}{2\tau_F}\right) - Si\left(\frac{t}{2\tau_F}\right)}{2\pi\tau_F} \right] \quad (33)$$

In this expression, we recognize the slow oscillation term while the fast oscillation term in Shastry's model has vanished because of the continuous character of Cattaneo's model.

In order to shed more light on the difference between the nonthermal contributions to the total energy density as described by Shastry's model and Cattaneo's model, we plot in Fig 4 the results of these two models in the case of gold at room temperature for comparison. Due to the band cut-off effect in Shastry's model, the nonthermal contribution shows exponentially damped oscillating behaviors as a function of time composed of two superimposed oscillations a fast and a slow one. On the other hand, because of the continuous character of Cattaneo's model, the fast oscillating behavior disappears and only remains the slow oscillations with very small amplitude. The nonthermal contribution shows almost an exponential decay as a function of time, much faster than the thermal contribution. As we can see in the inset of Fig 4, the nonthermal contribution to the total energy density in Cattaneo's model becomes almost insignificant after about $4\tau_F$-$6\tau_F$. This is faster than the time constant decay of the nonthermal contribution as calculated by Shastry's model.

We have mentioned earlier in the theory section, that the same time decomposition can be made in the frequency domain for both Shastry and Cattaneo models. In Fig 5, we show the frequency behaviors of the amplitude and phase of the nonthermal and the thermal contributions to the total energy density Green's function at the top free surface of gold and aluminum at room temperature, as calculated based on both models. The behavior of the total energy density is also reported for comparison. In Fig 6, we show only the result for gold at room temperature, to which we added Fourier's law prediction to check the behavior at low frequency regime.

As we can see in both Figs 5 and 6, the thermal contributions have similar behaviors in both models. On the other hand, the nonthermal contributions show different behaviors. In Cattaneo's model, the amplitude and phase are constant up to a certain frequency where their trends start to change. The amplitude decreases and stabilizes at a lower value at high frequencies, and the phase goes through a minimum and then comes back to the initial zero value. In Shastry's model, the nonthermal contribution shows a similar behavior to Cattaneo's model up to a certain frequency, after which, the behavior changes. The amplitude goes through a sharp maximum before it starts decaying at high frequencies, and the phase does not come back to the initial value, but it falls out rapidly to a value of $\pi/2$. These interesting additional features in the frequency behavior of the nonthermal contribution in Shastry's model are a consequence of the discrete character of the lattice.

Also we have reported in Fig 6(c) the behavior of the total energy density Green's function as a function of frequency as calculated according to Cattaneo, Shastry and Fourier models. We can see the perfect overlapping of the three models in the low frequency regime both for the amplitude and phase while in the high frequency regime the three models show different behaviors.

In a previous work [18], we presented the effect of temperature on the time behavior of both the thermal and nonthermal contributions to the energy density Green's function at the top free surface of a metal as calculated based on Shastry's model. We showed that, by decreasing the temperature, the amplitude of the thermal contribution decreases and flattens, on the other hand, the oscillating behavior is still the same and shows the same features, especially at short time scales where the fast oscillation dominates since the period of this latter oscillation $\theta_F$ is $\tau_F$ independent and as such temperature independent.

In Fig 7, we report the temperature effect on the frequency behaviors of the amplitude and phase of the thermal and nonthermal contributions to the total energy density Green's function at the top free surface of gold, as calculated based on both Cattaneo and Shastry models. Decreasing temperature has a systematic and straightforward effect. By decreasing the temperature, the total relaxation time $\tau_F$ increases and the characteristic features in each models shift to low frequency regime. The values of the total scattering relaxation time $\tau_F$ are calculated from the measured values of the electrical resistivity of gold at different temperatures [32] and the Fermi velocity is assumed to be constant. Table 2 summarizes the values of $\tau_F$ at different temperatures.

Let us get back to the time domain. So far we have made a full analysis of the time and frequency behavior of the nonthermal (ballistic) and thermal (ballistic-diffusive) contributions to the total energy density Green's function of electrons in the conduction band at the top free surface of a metal as calculated by both Shastry and Cattaneo models. We pointed out the similarities and differences between these two formalisms. In the following section we will see how the characteristic features in the energy transport, as predicted by these models, will change as we move the observation location in the direction of the energy transport through the medium far away from the excitation location. We will see that both terms which were called the "thermal" and "nonthermal" components in previous section, have actually a distinct wave-front giving a finite speed of propagation to the energy density. As we discussed previously, the nonthermal component was called as such, because it has oscillations due to the energy density that moves back and forth, and assigning a temperature to this part of energy does not really make sense.

We report in Figs 8(a) and 8(b) the temporal and spatial behaviors of the thermal contribution to the total energy density Green's function at different locations or different times and the result is compared with the Fourier's model. As we have concluded in the theory section, both Shastry and Cattaneo models predict the same behavior of the thermal contribution to the total energy density. The thermal contribution exhibits an energy pulse front shape that propagates with speed $U = (D_e/\tau_F)^{1/2}$ and whose amplitude decreases as one moves through the metal far from the excitation location. Because of the causality requirement, this contribution vanishes for location $y>\eta$. The locations beyond the energy pulse front remains unaltered, while at locations before the front, the behavior of the thermal contribution tends to a Fourier diffusive type law at long time scales as we can see in both Figs 8.

In Figs 9(a-d), the temporal and spatial behaviors of the nonthermal contribution to the energy density Green's function is reported for both Shastry and Cattaneo models. The nonthermal contribution exhibits a kind of a perturbed or distorted energy pulse front that propagates with the same speed U, rather than a smooth pulse front shape. In virtue of the causality requirements, the nonthermal contribution vanishes for locations $y>\eta$. The locations beyond the distorted energy pulse remain unaltered, while at locations before that, the nonthermal contribution manifests a damped oscillating behavior not only in time domain, but also in space. Indeed, by using the approximation we made earlier regarding the argument of $\overline{R_Q}$, meaning $\overline{R_Q} \approx 2Q$ for $Q \geq 1/2$, we can show that the nonthermal contribution to the total energy density in Shastry and Cattaneo models can be evaluated analytically using nondimensional variables $\eta$ and $y$. Based on Eq (10b), this gives for Shastry's model:

$$\delta K_S^>(\eta, y) = \frac{P_0}{\pi U \tau_F} e^{-\frac{\eta}{2}} \left\{ \begin{array}{l} \frac{Q_m}{2}\left[\sin c\left[Q_m(\eta+y)\right] + \sin c\left[Q_m(\eta-y)\right]\right] \\ -\frac{1}{4}\left[\sin c\left(\frac{\eta+y}{2}\right) + \sin c\left(\frac{\eta-y}{2}\right)\right] \\ +\frac{1}{4}\left[Si\left[Q_m(\eta+y)\right] + Si\left[Q_m(\eta-y)\right]\right] \\ -\frac{1}{4}\left[Si\left(\frac{\eta+y}{2}\right) + Si\left(\frac{\eta-y}{2}\right)\right] \end{array} \right\} H(\eta-|y|) \quad (34)$$

For Cattaneo's model, dropping off the first term in Eq (24b) and using the approximation on $\overline{R_Q}$ results on:

$$\delta K_S^>(\eta, y) = \frac{P_0}{\pi U \tau_F} e^{-\frac{\eta}{2}} \left\{ \begin{array}{l} \frac{\pi}{2}\left[\delta(\eta+y)+\delta(\eta-y)\right] \\ -\frac{1}{4}\left[\sin c\left(\frac{\eta+y}{2}\right)+\sin c\left(\frac{\eta-y}{2}\right)\right] \\ +\frac{\pi}{8}\left[1+Sgn(\eta-y)\right] \\ -\frac{1}{4}\left[Si\left(\frac{\eta+y}{2}\right)+Si\left(\frac{\eta-y}{2}\right)\right] \end{array} \right\} H(\eta-|y|) \quad (35)$$

Where *Sgn* refers to the *"sign function"* $Sgn(x) = +1$ if $x \geq 0$ and $-1$ if $x < 0$ [31].

Whether in Shastry's model or Cattaneo's model, both the temporal and the spatial behaviors of the nonthermal contribution to the total energy density are characterized by damped oscillations with the same periods; a fast and slow ones for Shastry's while only the slow oscillation remains in the case of Cattaneo's because of the continuous character of the latter. Slow oscillations are characterized by very small amplitude in comparison with fast oscillations. Generally, the amplitude of both oscillations decreases enormously at long times or farther distances and they become totally insignificant after about $\eta, y \geq 30$ as can be seen in the insets of Figs 9(a-d). The nondimensional fast and slow periods are given respectively by:

$$\begin{cases} \Theta_f^{Time} = \Theta_f^{Space} = 2\pi/Q_m = 2\sqrt{3}\frac{a}{v_F \tau_F} = 2\sqrt{3}\frac{a}{\Gamma_F} \\ \Theta_S^{Time} = \Theta_S^{Space} = 4\pi \end{cases} \quad (36)$$

Where, we have introduced the mean free path (MFP) of an electron in the conduction band of the metal $\Gamma_F = v_F \tau_F$.

The expressions of the periods in the nondimensional time-space domain can be related to their analogous in the dimensional time-space domain as:

$$\begin{cases} \Theta_f^{Time} = \dfrac{\theta_f^{Time}}{\tau_F} \text{ and } \Theta_f^{Space} = \dfrac{\theta_f^{Space}}{U\tau_F} \Rightarrow \theta_f^{Space} = 2a \\ \Theta_S^{Time} = \dfrac{\theta_S^{Time}}{\tau_F} \text{ and } \Theta_S^{Space} = \dfrac{\theta_S^{Space}}{U\tau_F} \Rightarrow \theta_S^{Space} = 4\pi U\tau_F = \dfrac{4\pi}{\sqrt{3}}\Gamma_F \\ \dfrac{\theta_f^{Space}}{\theta_f^{Time}} = \dfrac{\theta_S^{Space}}{\theta_S^{Time}} = U \end{cases} \quad (37)$$

The fast spatial period is nothing other than the size of the Wigner-Seitz unit cell, which is the reciprocal of the FBZ [23], while the slow spatial period is proportional to the MFP of the electron. Similar to the oscillations in the time domain, the oscillations in the space domain describe the same sequence of physical phenomena, namely the nonthermal (ballistic) transport of the energy density at short time scales when the distribution of the electronic system in the conduction band of the metal is still in a nonequilibrium state, followed by an oscillating transition to the thermal (ballistic-diffusive) regime as the electronic distribution tends towards an equilibrium thermal distribution and a temperature can be defined. A remarkable point in this analysis is the ratio of the spatial to the temporal periods of the fast and slow oscillations. This ratio is the same for both oscillations types and it is the speed of the energy pulse U.

Furthermore, based on the wave-vector separation value $q_0 = 1/2U\tau_F = \sqrt{3}/2\Gamma_F$ and the boundary of the FBZ $q_m = \pi/a$ in Shastry's model, we can define the range of the wavelengths of electrons that contribute to the nonthermal and thermal regimes in energy transport separately. It is straightforward to show that the wavelength of electrons that contribute to the nonthermal regime is between the fast and slow spatial oscillations periods $2a \leq \lambda_{NonTh}^e \leq \dfrac{4\pi}{\sqrt{3}}\Gamma_F \Leftrightarrow \theta_f^{Space} \leq \lambda_{NonTh}^e \leq \theta_S^{Space}$ and all electrons with a wavelength larger than the slow spatial oscillation period $\lambda_{Th}^e \geq \frac{4\pi}{\sqrt{3}}\Gamma_F \Leftrightarrow \lambda_{Th}^e \geq \theta_S^{Space}$ contribute to the thermal regime of energy transport. The separate contributions of electrons with different wavelengths are the result of neglecting electron-electron interactions in this analysis.

To end this section, we report in Figs 10 and 11, respectively, the temporal and spatial behaviors of the total energy density Green's function as calculated using Shastry, Cattaneo and Fourier models. For Cattaneo's model, we report both results of the conventional solution [Eq(16)], and the approximated solution [Eq(24)]. The total energy density undergoes similar behaviors as the thermal and nonthermal contribution separately, in each model. The amplitude decreases as the observation location is moved through the medium far from the excitation location and because of the causality requirement, the energy density vanishes for locations $y > \eta$. The locations beyond the energy pulse front remain unaltered, while at locations before the front, the energy density tends to a Fourier type behavior at long time scales as we can see in both Figs 10 and 11. In Figs 10(b) and 11(b), there is a slight discrepancy between the full conventional solution of Cattaneo's model and the approximated solution presented in this paper based on the decomposition of the total energy density Green's function to a nonthermal and a thermal contribution, even though both solutions show similar energy pulse front behavior. This slight discrepancy observed at short time scales tends to disappear at long time scales, and may come from the approximation we made on $\overline{R_Q}$ to calculate analytically the behavior of the nonthermal contribution. It should not however shadow the fundamental oscillating nature of the transition from the nonthermal regime to the thermal regime in energy transport predicted by this decomposition procedure that originates essentially from Shastry's model and that we extended to Cattaneo's model. On should note here, that because of the high amplitude of the oscillations, and in order to show how Shastry's model tend to Fourier's model at long time scales, we represented the vertical axis in Figs 10(a) and 11(a) in a logarithmic scale. This hides the energy front, but as we can see in the inset of Fig 10(a), Shastry's model shows a front just as Cattaneo's but with high amplitude oscillations superimposed.

One characteristic feature of the hyperbolic model of Cattaneo, which is reproduced by Shastry's model, is the competition between diffusion and accumulation of energy as the energy pulse moves through the medium far from the excitation location. This is illustrated in the insets of Figs 10(a) and 10(b). As one can see, when the observation location is far from the excitation location, the energy density tends to increase due to the accumulation effect before the diffusion catches on and starts to take over.

It may appear shocking or anomalous to have negative values of the energy density at short time scales, one should not forget however that at these short time scales, the electronic distribution in the conduction band of the metal is still in a nonequilibrium state and the energy density is mostly transported ballistically before it thermalizes. As the electronic distribution tends to an equilibrium thermal one, there is an exchange in energy density between different locations of the metal both ballistically and through diffusion (represented by $D_e$). This nonthermal (ballistic) transport of the energy density damps out with time very quickly as the nonthermal regime transition to a full diffusive thermal regime.

Many authors have reported the observation of an oscillating reflectivity change at the top free surface of semi-metals using the femtosecond pump-probe transient thermoreflectance technique [33-35]. Probing the relative change of the surface reflectivity is proportional to probing the change of the energy density at this surface. Nevertheless, these oscillations have been identified as due to a generation and relaxation of coherent phonon in such semi-metals. The frequencies of which have been confirmed using Raman Spectroscopy techniques [33-35]. By consequence, these oscillations are not related to the Bloch-type oscillation in the energy transport by the conduction band electrons in metals predicted by Shastry's formalism. Two main reasons can explain the lack of Shastry oscillations: *(i)* the smallness of the oscillation period *(1fs)* which make it impossible to observe unless using attosecond sources, *(ii)* the optical penetration depth of metals at long excitation laser wavelengths.

A possible candidate to observe the fast oscillations in the energy transport is a metallic superlattice with a high degree of coherence for electron transport through the interfaces. As suggested by the expression of the oscillation period $\theta_f$, this latter is proportional to the lattice constant, which is a consequence of the integration over the FBZ. It is well known however that superlattices structures are characterized by a subdivision of the electronic and phononic bands into mini-bands. Particularly the FBZ is divided into mini Brillouin zones of width $\pi/d$ where *d* is the superlattice period. This spatial period *d* can be one to two orders of magnitude larger than the lattice constant *a*, which will increase the energy density oscillating period by the same order of magnitude and will bring its value from the femtosecond regime to the picosecond regime. This latter regime can be probed by the state of the art in femtosecond laser metrology. As a matter of fact, the conventional Bloch oscillations have been only observed in superlattices structures [8, 9].

In addition, short pulse laser sources are in continuous development and attosecond width pulses have been developed [10]. Even though many other resonance phenomena of condensed matter have to be taken into account, these sources can be used to observe the fundamental energy transport oscillations. It is important to use short wavelengths, still longer than the Plasmon wavelength of the corresponding metal, in order to reduce the laser absorption distance.

# 4. SUMMARY

We have analyzed the transition between the nonthermal (ballistic) and the thermal (ballistic-diffusive) energy transport in metals using the recently developed Shastry's formalism. Interesting spatial and temporal oscillations in the energy density Green's function is reported under delta function excitation. Two types of oscillations are identified. The fast oscillation behavior in the energy transport is a consequence of the band cut-off due to the discrete character of the crystalline lattice. This leads to Bragg reflection of electrons in a metal. This fast oscillating behavior can be viewed as an energetic analogous to the conventional Bloch oscillation in the charge density of the conduction band electrons. It is an interesting manifestation of the ballistic and nonthermal contribution to the energy transport that results from the electrons bouncing back and forth at the boundaries of the first Brillouin zone before they damp out into the diffusive and thermal regime due to scattering mechanisms. On the other hand, the slow oscillation behavior describes the transition from the nonthermal regime to the thermal regime of energy transport. This transition does not occur in an abrupt way or gradually but rather in a damped oscillatory manner.

Remarkably, Cattaneo's model shows similar features as Shastry's formalism. More specifically, a similar decomposition in the total energy density of the metal in the time-space domain can be made. The thermal contribution to the energy density is described using a formula similar to Shastry's model. On the other hand, because of the continuous character of Cattaneo's model, the nonthermal contribution shows a different behavior. Cattaneo includes only the slow oscillations. The nonthermal contribution to the energy density appears to decay on a time constant even faster than the one predicted by Shastry's model.

In the frequency domain, the phase of the total energy density Green's function shows a $\pi/2$ shift at high frequencies. If the phase could be detected at very high frequencies, this difference can be probed and used as an indication of the additional oscillations in the time domain.

In this paper we did not consider electron-electron interactions since they don't change the total energy of the electron gas. However the redistribution of the electrons in the momentum space, could affect the ballistic distance travelled by electrons with different wavelengths. Since the electron-electron relaxation time could be on the order of the oscillation periods described earlier, a complete analysis should be based on the change in the electron distribution function more explicitly. While techniques such as Monte Carlo are very powerful, the complexity of individual scattering processes and the numerical accuracy make it difficult to detect wave fronts and the very fast oscillations. This is, however, a good direction for further study the transition between energy wave oscillations and the thermal transport.

The treatment in this paper for energy and heat transport by electrons in the conduction band of metals can easily be extended to semiconductors where the dominant energy and heat carriers are phonons. In this case, the total relaxation time in Shastry's model is wave-vector dependent and cannot be taken to be constant. The case of semiconductors is being investigated and will constitute the topic of a future work.

## ACKNOWLEDGMENTS

The authors would like to thank Professor B. S. Shastry for his enlightening discussions throughout this work. This work was supported by the Interconnect Focus Center, a DARPA and Semiconductor Research Corporation program, as well as the AFOSR Thermal Transport MURI.

**List of table captions**

**Table 1:** Properties of gold and aluminum used in the calculations at room temperature.

**Table 2:** Properties of gold used in the calculations at different ambient temperatures.

**List of figure captions**

**Figure 1:** (Color online) Schematic diagram of the metal being excited by a laser delta pulse at its free top surface (x=0).

**Figure 2:** (Color online) Temporal behavior of the thermal (Eq (8a), (a)) and nonthermal (Eq (8b), (b)) parts of the energy density Green's function in Shastry's model at the top free surface of gold and aluminum at room temperature. The inset in (b) shows a zoom of the behavior of nonthermal contribution at long time scale for gold.

**Figure 3:** (Color online) (a) Comparison between the thermal contribution to the total energy density Green's function at the top free surface of gold and aluminum with Fourier's model at room temperature. (b) Comparison between the thermal contribution (dashed line), the sum of the nonthermal and thermal contributions (solid line) to the total energy density Green's function at the top free surface of gold at room temperature, with Fourier's model (dotted line).

**Figure 4:** (Color online) Comparison between the temporal behaviors of the nonthermal contribution to the total energy density Green's function at the top free surface of gold at room temperature, as calculated based on Shastry's model (solid line) and Cattaneo's model (dashed line in the inset).

**Figure 5:** (Color online) Frequency behavior of the thermal contribution (a, b), the nonthermal contribution (c, d) and the total energy density Green's function (e, f) at the top free surface of gold and aluminum at room temperature in both Cattaneo's model (a-c) and Shastry's model (d-f).

**Figure 6:** (Color online) Frequency behavior of the thermal contribution (a), the nonthermal contribution (b) and the total energy density Green's function (c) at the top free surface of gold at room temperature in both Cattaneo's model (solid line) and Shastry's model (solid-dashed line). The dashed line in (c) describes Fourier's model.

**Figure 7:** (Color online) Frequency behavior of the thermal contribution (a, b), the nonthermal contribution (c, d) and the total energy density Green's function (e, f) at the top free surface of gold at different temperatures in both Cattaneo's model (a-c) and Shastry's model (d-f).

**Figure 8:** (Color online) Comparison between the behaviors of the thermal contribution to the energy density Green's function of gold at room temperature as calculated based on both Shastry and Cattaneo models (solid line) with Fourier's model (dashed line), (a) temporal behavior at different locations $y$ and (b) spatial behavior at different moments $\eta$.

**Figure 9:** (Color online) Temporal (a, b) and spatial (c, d) behaviors of the nonthermal contribution to the energy density Green's function of gold at room temperature as calculated using Shastry's model (a, c) and Cattaneo's model (b, d). The insets in (a) and (b) show a zoom of the slow oscillations while the insets in (c) and (d) show the spatial behavior of the nonthermal contribution at $\eta=30$.

**Figure 10:** (Color online) (a) Temporal behavior of the total energy density Green's function of gold at room temperature at different locations $y$, as calculated based on Shastry's model (solid line), in comparison with full Cattaneo's model (dashed line) and Fourier's model (dotted line). (b) Temporal behavior of the total energy density Green's function of gold at room temperature at different locations $y$, as calculated based on approximated Cattaneo's model (solid line), in comparison with full Cattaneo's model (dashed line) and Fourier's model (dotted line).

**Figure 11:** (Color online) (a) Spatial behavior of the total energy density Green's function of gold at room temperature at different times $\eta$, as calculated based on Shastry's model (solid line), in comparison with full Cattaneo's model (dashed line) and Fourier's model (dotted line). (b) Spatial behavior of the total energy density Green's function of gold at room temperature at different times $\eta$, as calculated based on approximated Cattaneo's model (solid line), in comparison with full Cattaneo's model (dashed line) and Fourier's model (dotted line).

| Metal | Gold | Aluminum |
|---|---|---|
| **Lattice constant (Å)** | 4.08 | 4.05 |
| **Relaxation time** $\tau_F$ **(fs)** | 28 | 5.2 |
| **Fermi velocity** $v_F$ **($10^6$ m/s)** | 1.4 | 2.03 |

**Table 1:**

| Temperature T (K) | Electrical resistivity $\rho$ ($10^{-8}$ Ω.m) | Relaxation time $\tau_F$ (fs) |
|---|---|---|
| 273 | 2.04 | 30 |
| 169 | 0.592×2.04 | 50 |
| 90 | 0.270×2.04 | 109 |
| 68 | 0.177×2.04 | 167 |

**Table 2:**

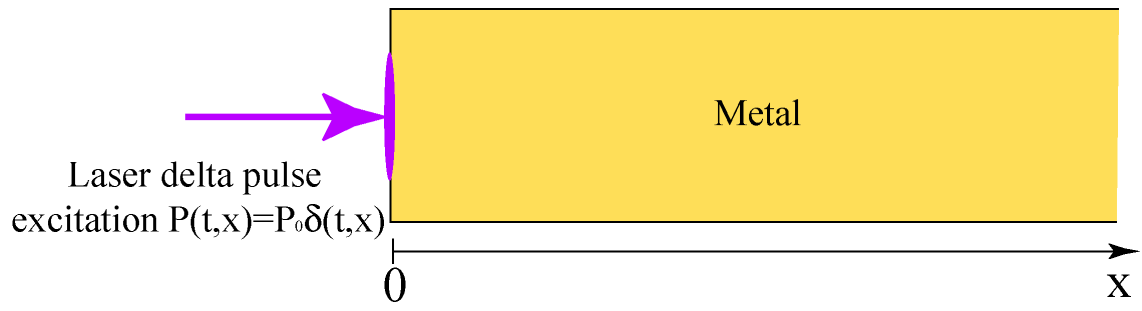

**Figure 1:**

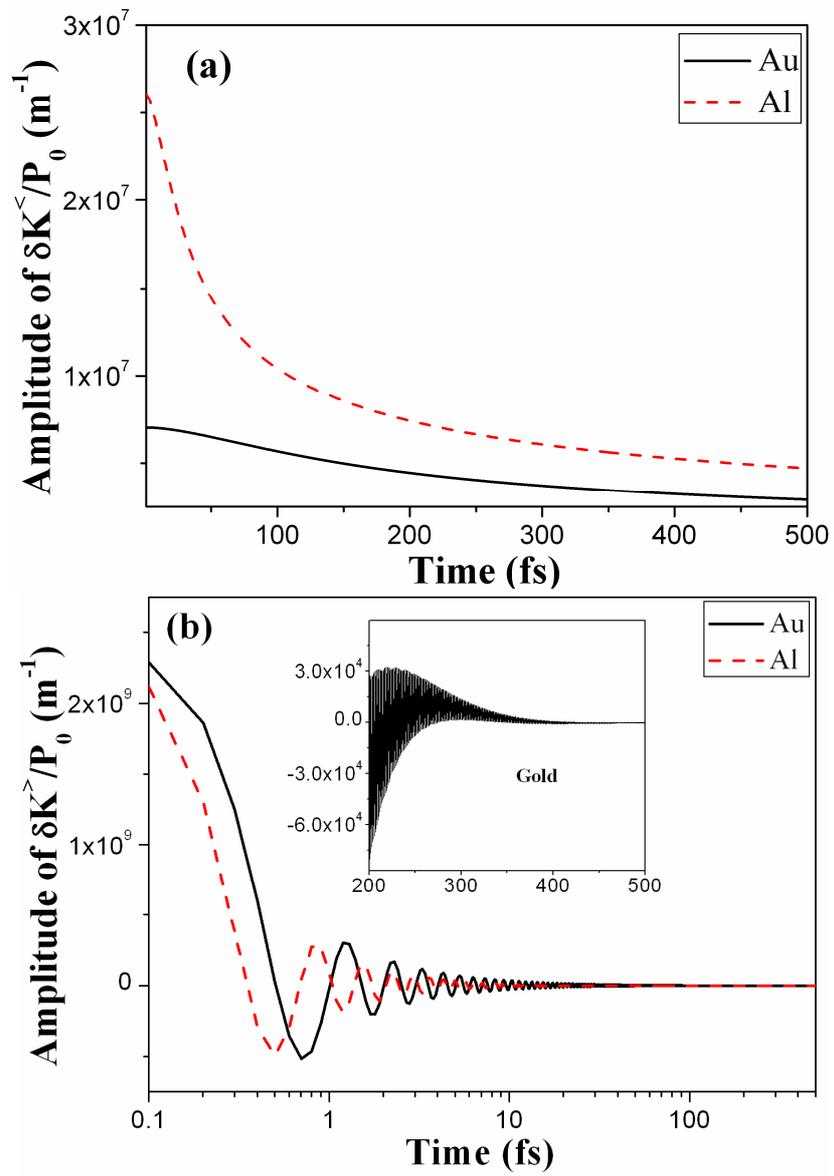

**Figure 2:**

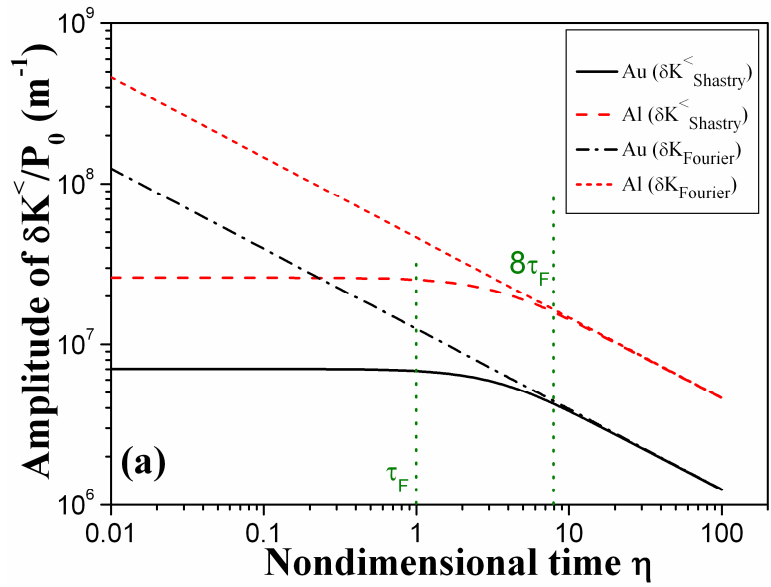

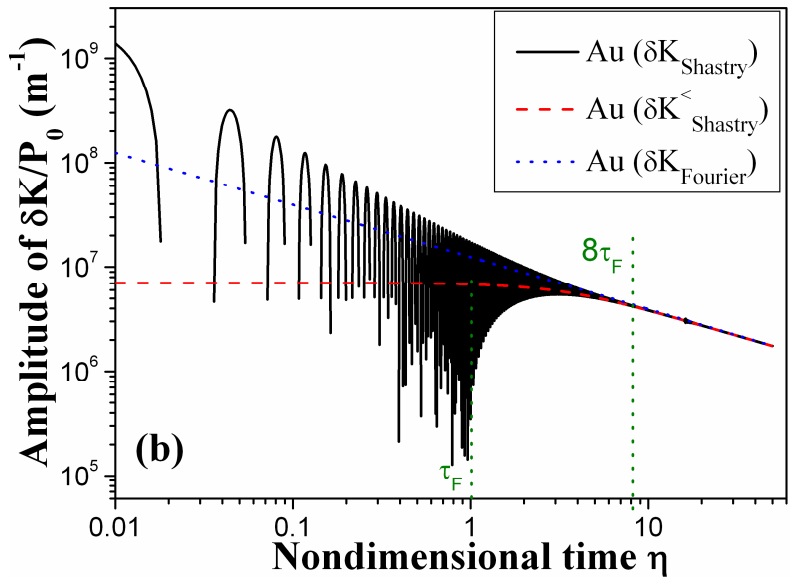

**Figure 3:**

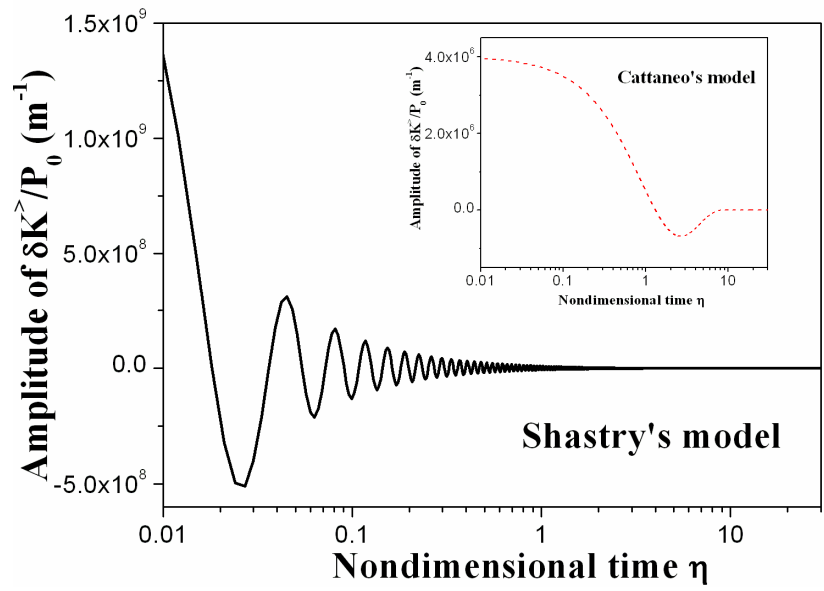

**Figure 4:**

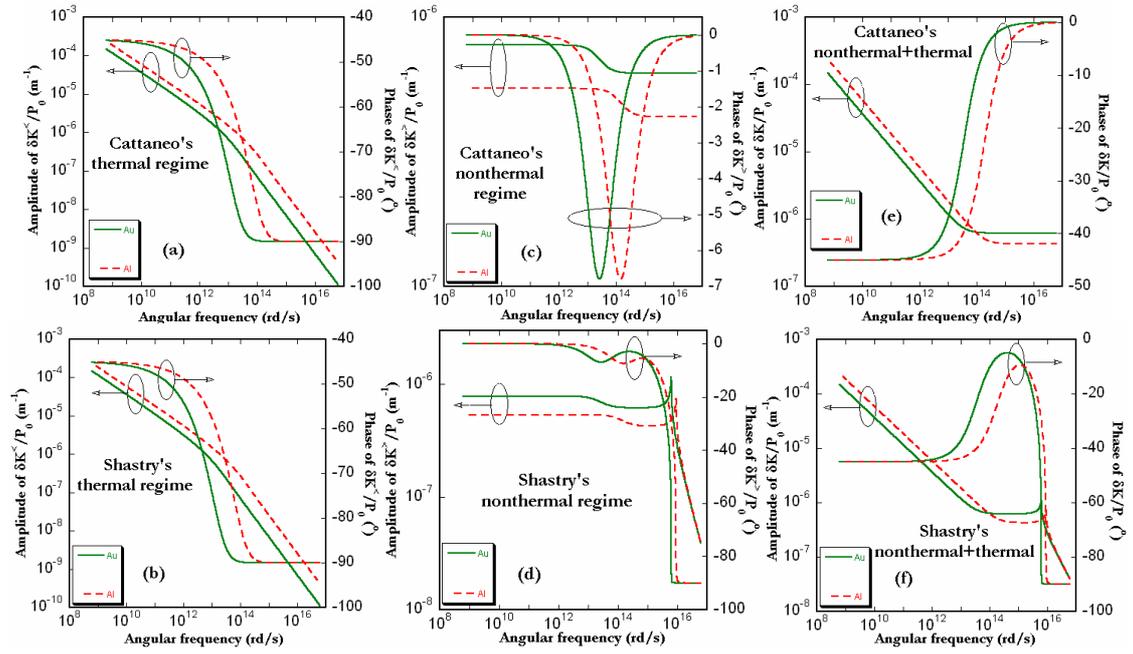

**Figure 5:**

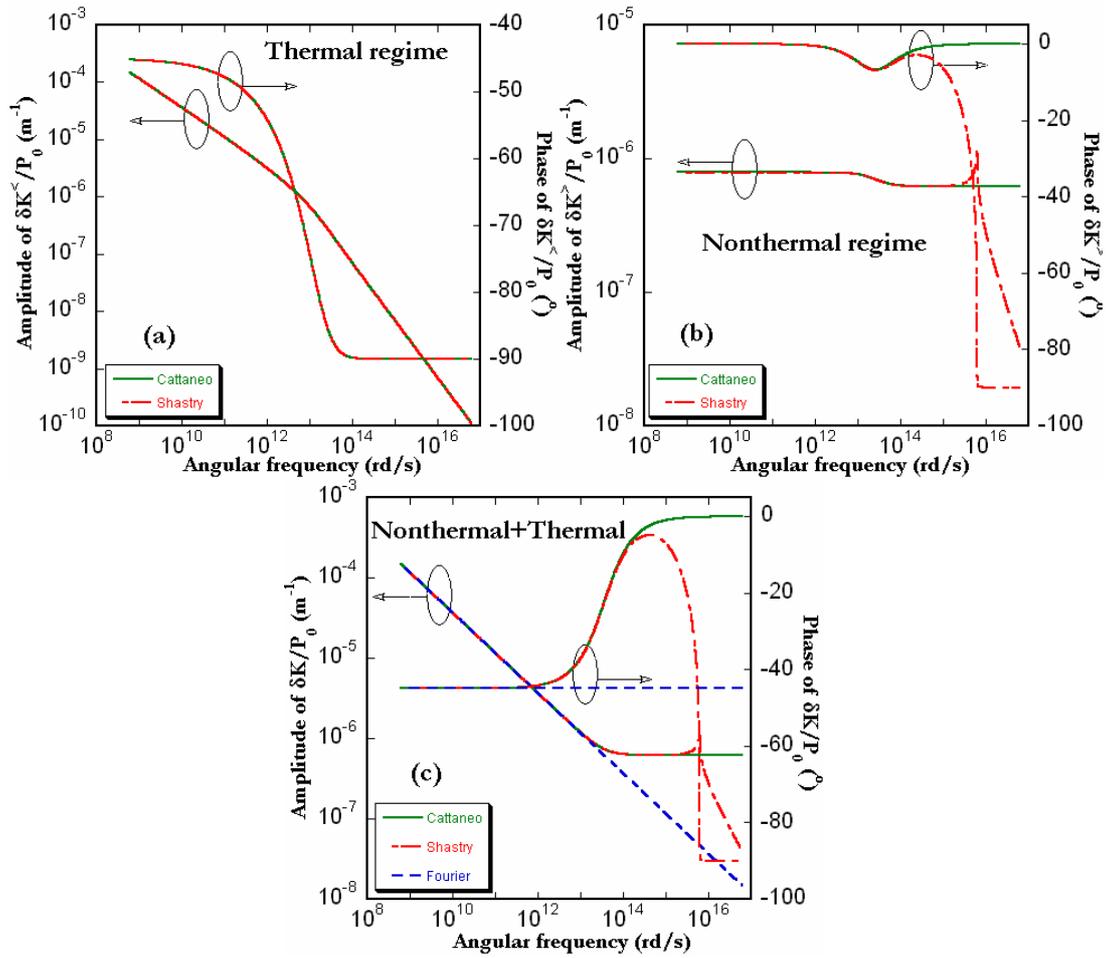

**Figure 6:**

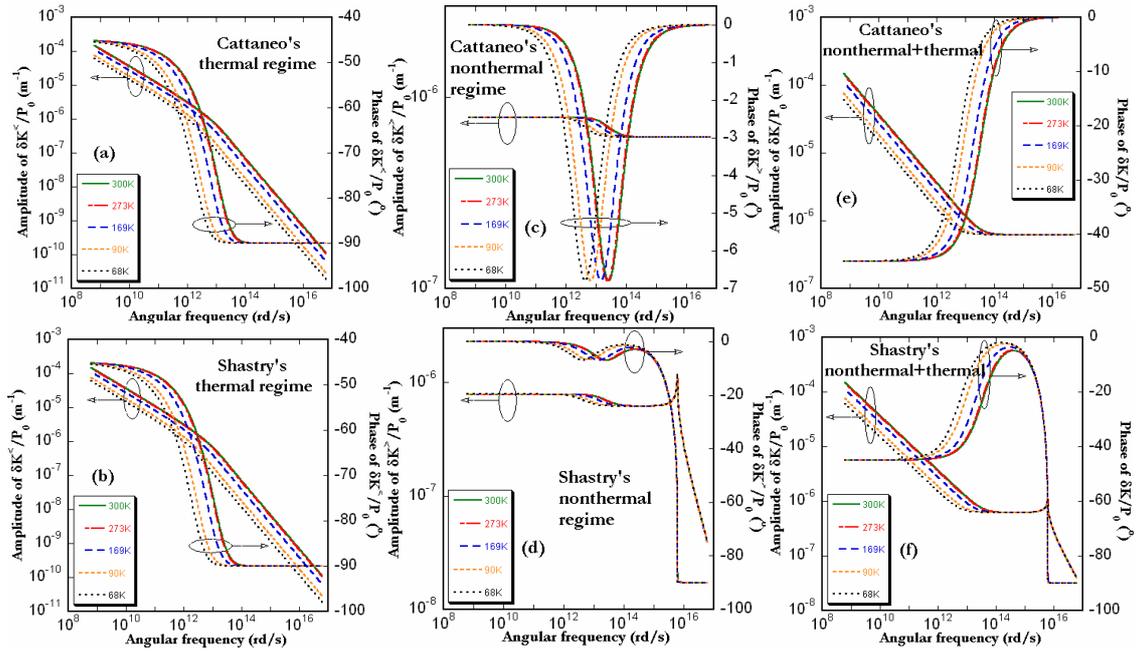

**Figure 7:**

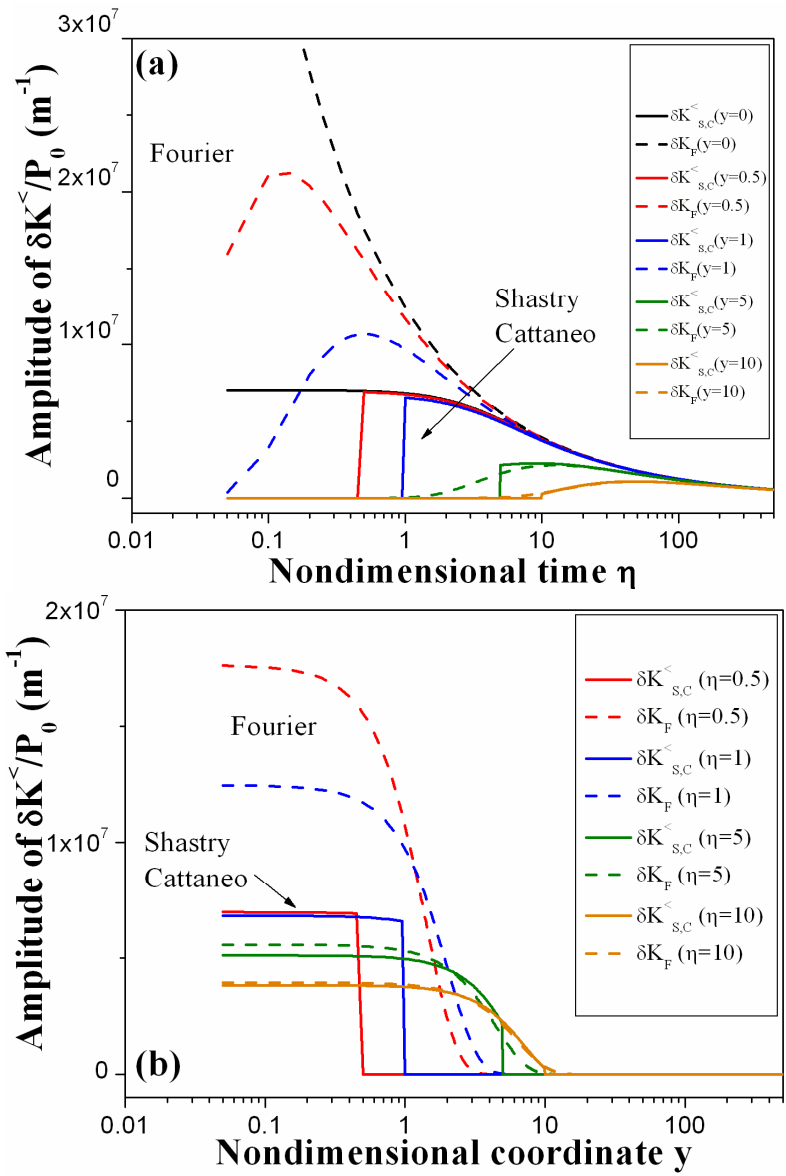

**Figure 8:**

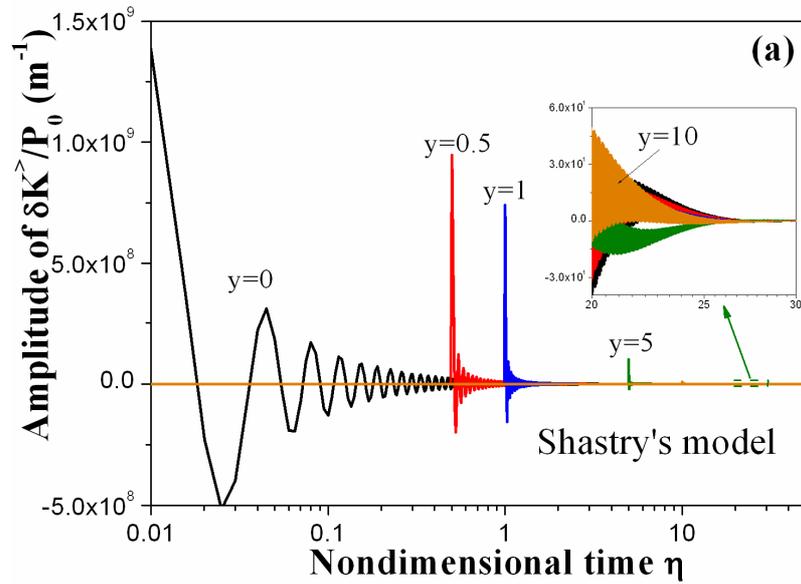

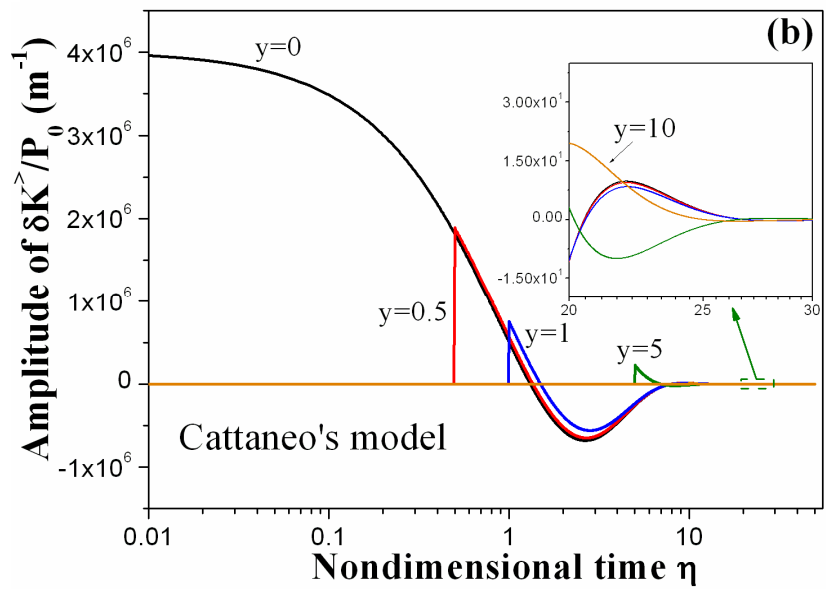

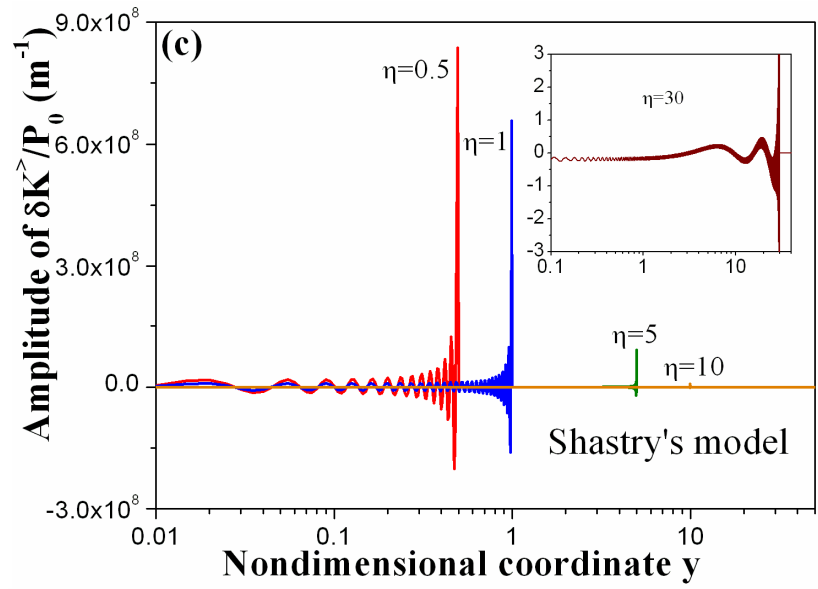

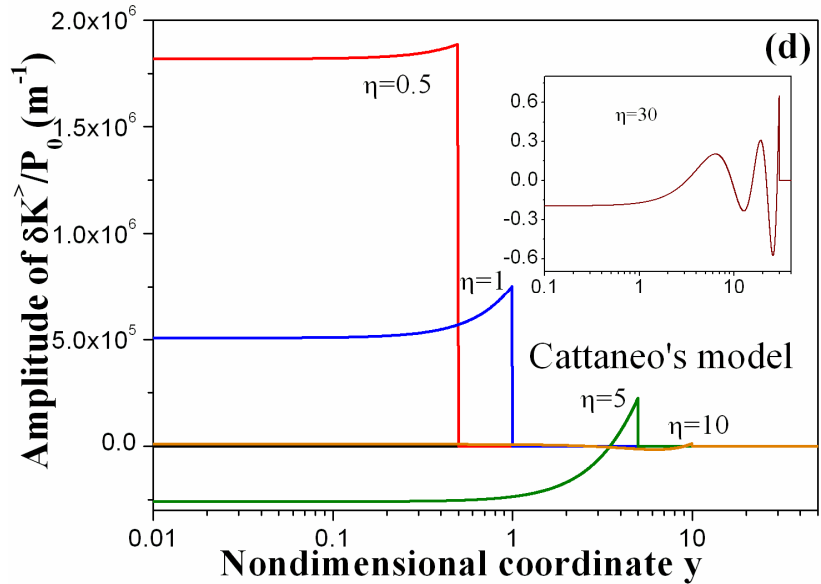

**Figure 9:**

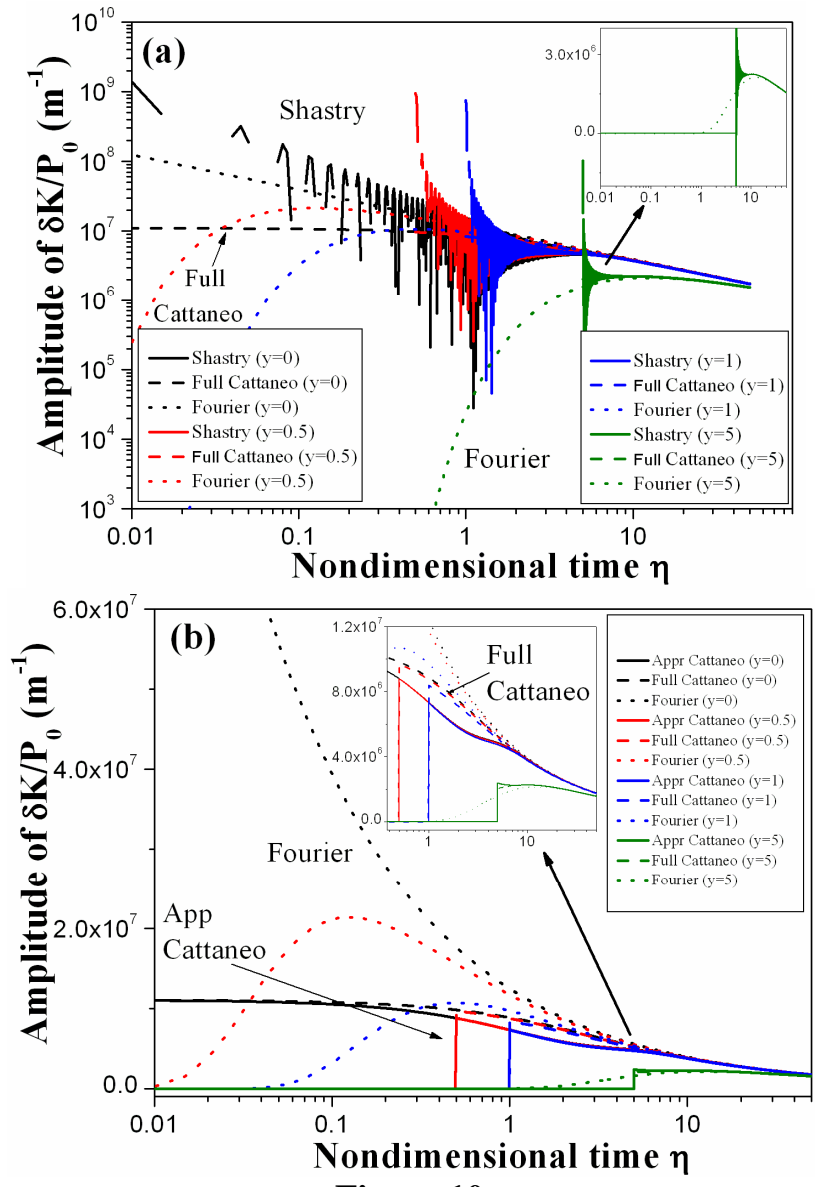

Figure 10:

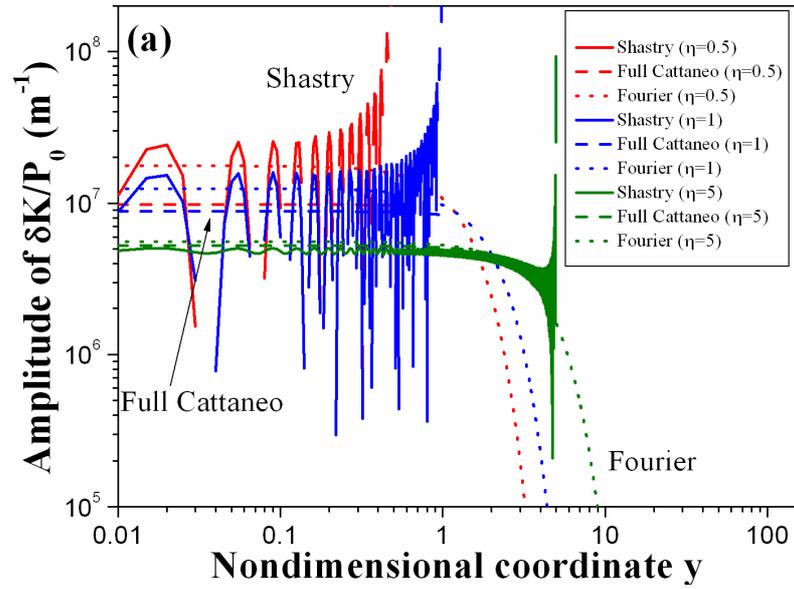
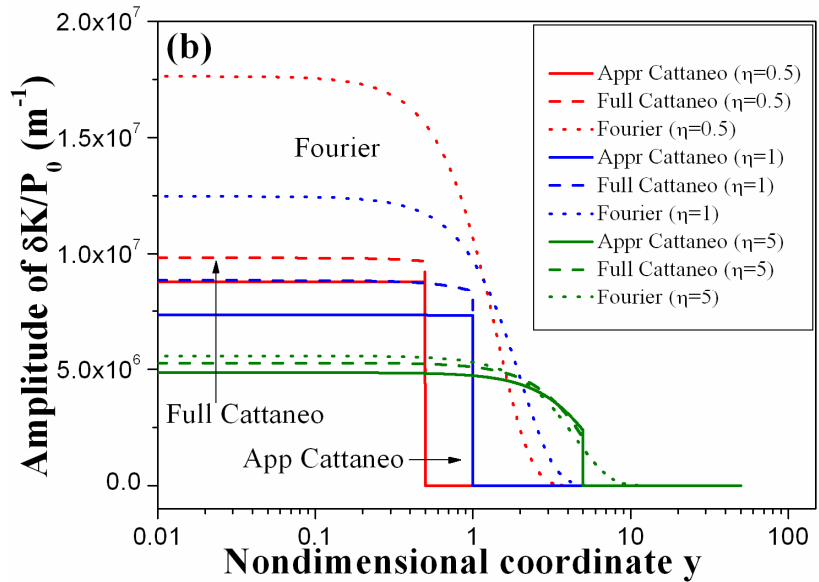

**Figure 11:**